\documentstyle[12pt]{article}
\title{Differential equations for \\
definition and evaluation of Feynman integrals}

\author{F. A. Lunev \thanks{e-mail address:
lunev@hep.phys.msu.su } \\ {\em Physical Department, Moscow
State University, Moscow, 119899, Russia} }

\date{ \ \ \  }

\begin{document}
\baselineskip 24pt

\maketitle

\begin{abstract}
It is shown that every Feynman integral can be interpreted as
Green function of some linear differential operator with
constant coefficients. This definition is equivalent to usual
one but needs no regularization and application of $R$-operation.
It is argued that presented formalism is convenient for practical
calculations of Feynman integrals.
\end{abstract}

\vspace{1.5cm}

Though  fundamental   results  in   renormalization  theory   were
obtained    many    years    ago    in    classical    works   of
Feynman,   Tomonaga,   Schwinger,    Dyson,   Salam,    Bogolubov,
Parasiuk,  Hepp,   and   Zimmermann\footnote{Beautiful account of
foundations and modern achievements of renormalization theory can
be found, for instance, in monographs \cite{BS,Col,Sm}},
renormalization problems continue to attract the attention of
theorists.   In particular, during last twenty years   very many
papers were devoted to investigations of various regularization
schemes.

Of course,  all known  regularization schemes  are equivalent,  in
principal, at perturbative level. However, their practical  value
is different.  For instance,  only the  discovery of  dimensional
regularization  \cite{DimR} gave  possibility  to  carry  out
systematical calculations    in    gauge    theories.
Moreover,   different regularization  schemes,  that  are
equivalent  on   perturbative level,  can  be  nonequivalent
beyond  perturbation  theory.  For instance,  partial  summing
of  perturbation  series by means of renormalization  group
methods can give scheme dependent  results (see, for instance,
\cite{Sh}    .)    This    fact    stimulates further   search
of   "the   most   natural"   and   convenient regularization
scheme.

In  this  paper  we  will  show  that  Feynman  integrals  can be
defined and  evaluated without   any regularization  at   all. Of
course,   in  itself  it   is  not   a  surprise.  In particular,
recently proposed  differential     regularization      \cite{DR}
also      needs      no  regularization  in  usual  sense.    But
simplicity of our  results is  the real  surprise. {\em   We will
show that   any Feynman  diagram without  internal   vertexes can
be  treated   as  Green   function  of  some linear  differential
operator  with constant  coefficients.} This result allows   also
to define  and evaluate Feynman  diagram with internal  vertexes,
because   such diagrams  can be   considered as  certain diagrams
without  internal  vertexes   at  zero  value   of  some external
momenta.   For instance,  the   value of  diagram   with internal
vertexes   on Fig.1   coincides with   the value   at $k=q=0$   of
diagram on Fig.2 .

\begin{figure}[p]
\unitlength=5mm
\thicklines
\begin{picture}(12,10)(-10,0)
\put(5,5){\oval(4,4)[l]}
\put(7,5){\oval(4,4)[r]}
\put(5,3){\line(1,0){2}}
\put(6,7){\circle{2}}
\put(1,5){\vector(1,0){1}}
\put(2,5){\line(1,0){1}}
\put(9,5){\vector(1,0){1}}
\put(10,5){\line(1,0){1}}
\put(2,6){p}
\end{picture}
\caption{Example of a diagram with internal vertexes}
\end{figure}
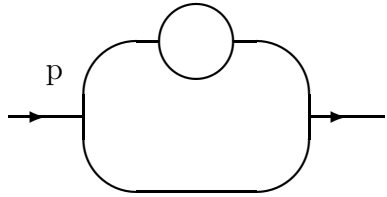

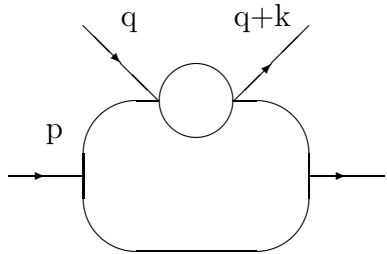
\begin{figure}[p]
\unitlength=5mm
\begin{picture}(12,10)(-10,0)
\put(5,5){\oval(4,4)[l]}
\put(7,5){\oval(4,4)[r]}
\put(5,3){\line(1,0){2}}
\put(6,7){\circle{2}}
\put(1,5){\vector(1,0){1}}
\put(2,5){\line(1,0){1}}
\put(9,5){\vector(1,0){1}}
\put(10,5){\line(1,0){1}}
\put(2,6){p}
\put(5,7){\line(-1,1){1}}
\put(7,7){\vector(1,1){1}}
\put(3,9){\vector(1,-1){1}}
\put(8,8){\line(1,1){1}}
\put(4,9){q}
\put(7,9){q+k}
\end{picture}
\caption{The diagram without internal vertexes that corresponds
to one on Fig.1}
\end{figure}

Renormalization  scheme,  given  in  this  paper,   is
equivalent to usual R-operation  scheme.But "equivalent"
doesn't mean "the same". Indeed, in standard  R-operation
renormalization  scheme  one  must,  first,  regularize  initial
divergent (in  general) Feynman  integral.Then it is necessary
to use rather complicated subtraction  prescription (forest
formula) to obtain finite   result.    Nothing    similar   is
needed in    my renormalization  scheme.  To  obtain  finite
expression for given Feynman  integral,  one  must   only  solve
some well   defined differential  equations.  Neither  any
regularization,  nor  any manipulations  with  counterterm
diagrams  are  needed to obtain finite result.

For  simplicity,  in  this  paper we will consider
only   scalar    Feynman   integrals.   General   case   will  be
investigated in more long forthcoming paper.

Let  us  consider  arbitrary  (Euclidean) Feynman diagram without
internal  vertexes  in  coordinate  space.  This  is well defined
function

\equation
{\tilde {\Gamma}}_n (x_1,...,x_n; \{ m_{ij}^2 \}) = \prod_{all \
lines \ of \ \Gamma} D(x_i - x_j; m_{ij}^2) \label{1}
\endequation

\noindent where

$$D(x,m^2) = \int \mbox{d}^4 p \frac{e^{ipx}}{p^2+m^2}  $$

\noindent But their Fourier image

\equation
\Gamma _n (p_1,...,p_{n-1}; \{ m_{ij}^2 \} ) =  \frac{1}{{(2 \pi
^4)}^{n-1}} \int \mbox{d}^4 x _1...\mbox{d}^4 x_{n-1} \exp(ip_1
x_1 + ... + ip_{n-1} x_{n-1}) \tilde {\Gamma}
(x_1,...,x_{n-1},0; \{ m_{ij}^2 \} ) \label{1a}
\endequation

\noindent is not, in general, well defined. The problem of
renormalization theory is to define the function $\Gamma
(p_1,...,p_{n-1}; \{ m_{ij}^2 \})$ .

Below we will interpret $ m_{ij}^2 $ as the square of some
Euclidean two dimensional vector. Then we can write

$$m_{ij}^2 = (m_{ij,1})^2 + (m_{ij,2})^2 $$

\noindent Further, let us define the Fourier image of $
D(x,m^2)$ with respect to variables $m_1$ and $m_2$ (here $m_1^2
+ m_2^2 = m^2$):

$$\hat D (x,u) = \int \mbox{d}^2 m e^{i \vec u \vec m} D(x,m)=
\int \mbox{d}^4 p \mbox{d}^2 m \frac{e^{ipx + i \vec u \vec
m}}{p^2 +m^2} $$

\noindent   where   $\vec   u    =(u_1,u_2),   \   u^2  =   u_1^2
+  u_2^2  $.  It  follows  from  definition that $\hat D(x,u)$ is
the Green    function   of    six dimensional Laplace operator

$$ \bigtriangleup _{xu} = \sum_{i=1}^4 \frac{\partial ^2}{\partial
x_i^2}+
\sum_{i=1}^2 \frac{\partial ^2}{\partial u_i^2}  $$

So

$$ \hat D (x,u) = 16 \pi ^3 \frac{1}{(x^2 + u^2)^2} $$

\noindent and

$$ \hat {\Gamma} _n (x_1,...,x_n; \{ u_{ij}^2 \}) \equiv \int
\prod _{all \ lines \ of \ \Gamma } \mbox{d} ^2 m_{ij} \exp
\left( i \sum _{all \ lines \ of \ \Gamma } \vec m _{ij} \vec u
_{ij} \right) \tilde \Gamma _n (x_1,...,x_n; \{m_{ij}^2 \} ) $$
$$=(16 \pi ^3)^N \frac{1}{P(x_1,...,x_n; \{ u_{ij}^2 \} ) }$$

\noindent where $N$ is total number of lines in diagram $\Gamma$
and $P$ is the polynomial:
\equation
  P = \prod _{all \ lines \ of \ \Gamma } [(x_i - x_j)^2 +
u_{ij}^2]^2 \label{2a}
\endequation

We see that $\hat \Gamma $ satisfies simple algebraic equation

\equation
P(x_1,...,x_n; \{ u_{ij}^2 \} ) \/ \hat {\Gamma} (x_1,...,x_n;\/
\{ u_{ij} \} ) = (16 \pi ^3)^N  \label{2b}
\endequation

Comparing (\ref{1a}), (\ref{2a}) and (\ref{2b}), we see that it
is very naturally {\em to define } $ \Gamma (p_1,...,p_{n-1}) $ as a
solution of differential equation

\eqnarray
P \left(i \frac {\partial}{\partial p_1},...,i \frac
{\partial}{\partial p_{n-1}},0;
\{\bigtriangleup _{m_{ij}}\} \right) \Gamma _n (p_1,...,p_{n-1}; \{
m_{ij}^2 \}) \nonumber \\
= (16 \pi ^3)^N \delta (p_1)...\delta (p_{n-1}) \prod
_{all \ lines \ of \ \Gamma}  \delta ( \vec {m_{ij}} ) \label{3}
\endeqnarray

\noindent This means that $\Gamma $ is Green function of linear
differential operator
$P \left(i \partial / \partial p_1 ,...
 \right) $. For instance, the diagram on Fig.2
is defined by equation

$$ (\bigtriangleup _{pm_1} )^2 \/(\bigtriangleup _{(p-q)m_2} )^2
\/ (\bigtriangleup _{km_3} )^2 \/ (\bigtriangleup _{qm_4} )^2
\/(\bigtriangleup _{qm_5} )^2 \/ \Gamma = (16 \pi ^3)^5 \delta
(p) \delta (q) \delta (k) \prod _{i=1}^{5} \delta (\vec {m_i} )
$$

\noindent where

$$ \bigtriangleup _{(p-q)m_2} = \sum _{i=1}^{4} \left( \frac
{\partial}{\partial p_i} - \frac {\partial}{\partial q_i}
\right)^2 + \sum _{i=1}^{2} \frac { {\partial}^2}{{\partial
m_{2,i}}^2} $$

Eq.(\ref{3}) defines $\Gamma$ up to solution of homogeneous
equation

$$P \left(i \frac {\partial}{\partial p_1},...,i \frac
{\partial} {\partial p_{n-1}},0; \{\bigtriangleup _{m_{ij}}\}
\right) \Gamma = 0 $$

\noindent This arbitrariness can be fixed inductively in the
following way.

Let all diagrams with ($L-1$) loops are already defined. For
given $L$-loop diagram with divergent index $\omega (\Gamma )$
one defines $(L-1)$-loop diagram $\Gamma _{ij}$ as diagram
$\Gamma$ without the line $(ij)$ with propagator $((p - k)^2 +
m_{ij}^2)^{-1}$ where $p$ is external momentum. We can always
define external momenta in such way that $\Gamma _{ij}$ doesn't
depend on $p$. (See Fig.3, where $\Gamma _{ij}$ is represented
as shaded block.)

\begin{figure}[p]
\unitlength=0.1mm
\centering
\begin{picture}(400,400)
\thicklines
\put(200,200){\oval(200,100)}
\put(200,250){\oval(100,100)[t]}
\put(50,350){\vector(1,-1){50}}
\put(250,250){\vector(1,1){50}}
\put(100,300){\line(1,-1){50}}
\put(300,300){\line(1,1){50}}
\put(150,150){\line(-1,-1){100}}
\put(250,150){\line(1,-1){100}}
\put(150,75){\circle*{8}}
\put(200,75){\circle*{8}}
\put(250,75){\circle*{8}}
\put(50,300){p}
\put(350,300){p+q}
\thinlines
\put(100,200){\line(1,1){50}}
\put(200,250){\line(-1,-1){86}}
\put(250,250){\line(-1,-1){100}}
\put(200,150){\line(1,1){86}}
\put(250,150){\line(1,1){50}}
\end{picture}
\caption{Illustration to the proof of equivalence of proposed
renormalization scheme and usual one}
\end{figure}
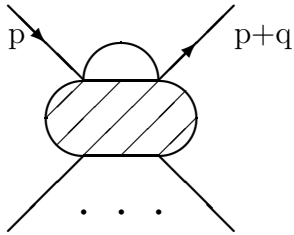

It is easy to see that if $\Gamma $ satisfies the equations

\equation
(\bigtriangleup _{pm_{ij}})^2 \Gamma = (16 \pi ^3) \/ \delta (\vec
m_{ij})
\Gamma _{ij} \label{5}
\endequation

\noindent then $\Gamma$ also satisfies the equation (\ref{3}).
Finally, we impose asymptotic conditions

\equation
\lim_{|p | \rightarrow \infty}  \frac {1}{p^{\omega (\Gamma ) +
\epsilon}} \Gamma = 0; \ \
\lim_{m_{ij}  \rightarrow \infty}  \frac {1}{m^{\omega (\Gamma
) + \epsilon}_{ij}} \Gamma = 0;  \label{6}
\endequation

\noindent for any $ \epsilon > 0 $.

It  is  easy  to  prove  that  equations  (\ref{5}) together with
analogous equations  for other  lines with  asymptotic conditions
(\ref{6}) define  $\Gamma$ up  to polynomial  of degree  $ \omega
(\Gamma ) $ with respect to external momenta and masses.  (For  $
\omega  (\Gamma  )  <  0  $  the  diagram  $\Gamma$   is  defined
unequivocally if $(L-1)$-loop diagrams are already defined.) So
our definition reproduces usual renormalization arbitrariness in
the definition of Feynman integrals.

Now  let  us   prove  that  for   renormalizable  theories   with
divergent index less  or equal two  our definition is  equivalent
to usual one.  Consider  again the diagram on   Fig.3 regularized
by  means  of  the   cut  off  at   large  momentum  $  \Lambda$.
We  will  denote  this  diagram  as  $\Gamma  _{\Lambda}$.    The
renormalized diagram  $\Gamma ^{ren}  (\Lambda)$ is  the sum   of
$\Gamma  _{\Lambda}$  and  counterterm  diagrams.  The later ones
can  be  divided   in   two  sets.   First  set   of  counterterm
diagrams  contains   the   line  $(ij)$.   The  sum  of   $\Gamma
_{\Lambda}$ and these diagrams can be written as

\equation
\int _{|k| < \Lambda } \mbox{d} ^4 k \/ \frac {1}{ (p - k)^2
+ m_{ij}^2 } \Gamma ^{ren}_{ij} (\Lambda)   \label{7}
\endequation

\noindent where $\Gamma ^{ren}_{ij} (\Lambda)$  is renormalized
diagram $\Gamma _{ij}$. This diagram doesn't depend on $p$.

The  second  set  of counterterm diagrams  doesn't  contain the
line $(ij)$.  They  are  produced  by  change  of some divergent
subdiagrams of $\Gamma$, that contain the line  $(ij)$ , on
polynomials not  more then  the  second  degree  with  respect
to external momenta and masses (See, for instance
\cite{BS,Col,Sm}).  In particular, these polynomials are not
more then the  second degree with respect to $p$  and $m_{ij}$.
The diagram $\Gamma ^{ren} (\Lambda)$ is  the sum of  (\ref{7})
and these polynomials.

Using the formulae

\equation
(\bigtriangleup _{pm_{ij}})^2 \frac{1}{(p-k)^2 + m_{ij}^2} =16 \pi ^3
\delta (p-k) \delta (\vec m _{ij})   \label{7a}
\endequation

\noindent and (\ref{7}), one can prove that

\equation (\bigtriangleup _{pm_{ij}})^2 \Gamma ^{ren} (\Lambda)
= 16 \pi ^3 \delta ({\vec m}_{ij}) \theta (\Lambda - |p_i|)
\Gamma ^{ren}_{ij} (\Lambda) \endequation

\noindent (because above mentioned polynomials are annihilated by
$(\bigtriangleup _{pm_{ij}})^2 $.) In the limit $\Lambda
\rightarrow \infty $ one obtains the equation (\ref{5}).
Asymptotic conditions (6) are satisfied due to Weinberg's
theorem \cite{W} . This finishes the prove.

Now let us consider an illustrative example that show how our
definition works in  practical calculations.

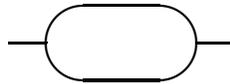
\begin{figure}[p]
\unitlength=5mm
\centering
\thicklines
\begin{picture}(8,5)
\put(1,4){\line(1,0){1}}
\put(4,4){\oval(4,2)}
\put(6,4){\line(1,0){1}}
\end{picture}
\caption{The simplest divergent Feynman diagram}
\end{figure}

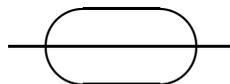
\begin{figure}[p]
\unitlength=5mm
\centering
\thicklines
\begin{picture}(8,5)
\put(1,4){\line(1,0){6}}
\put(4,4){\oval(4,2)}
\put(6,4){\line(1,0){1}}
\end{picture}
\caption{"Raising sun" diagram}
\end{figure}

\vspace{0.8cm}

The diagram $\Gamma ^{(1)}$ on Fig.4 is the simplest divergent
Feynman one. In according to our general theory it is defined
by equations

\eqnarray
(\bigtriangleup _{pm_1})^2 \Gamma ^{(1)} = \frac{16 \pi ^3
\delta (\vec m _1)}{p^2         +         m_2^2}
\label{7b}          \\ (\bigtriangleup _{pm_2})^2 \Gamma ^{(1)}
= \frac{16 \pi ^3 \delta (\vec m _2)}{p^2             +
m_1^2}              \label{8} \endeqnarray

Using the formula \footnote{Representation (\ref{Mel}) was used
first for evaluation of Feynman integrals in \cite{BD}}

\equation
 \frac{1}{ {p^2 + m_2^2}} =
 \frac{1}{2 \/ \pi i } \int \limits
_{\begin{array}{c}
\scriptstyle
c - i \infty \\
\scriptstyle
0<Re \ c< 1
\end{array} }
^{c + i \infty}
\mbox{d} s \Gamma (s) \Gamma (1-s) \frac
{(m^2_2)^{s-1}}{(p^2)^s}
\label{Mel}
\endequation

\noindent one can represent the solution of (\ref{7b}) in the
form

\equation
\Gamma ^{(1)}=
 \frac{1}{2 \/ \pi i } \int \limits
_{\begin{array}{c}
\scriptstyle
c - i \infty \\
\scriptstyle
0<Re \ c< 1
\end{array} }
^{c + i \infty}
\mbox{d} s \Gamma (s) \Gamma (1-s) (m^2_2)^{s-1} \Gamma _s +
f_1(m^2_2)   \label{9}
\endequation

\noindent where $\Gamma _s$ satisfies  the equation

\equation
(\bigtriangleup _{pm_1})^2 \Gamma _s = \frac{16 \pi ^3 \delta (\vec m
_1)}{(p^2)^s} \label{10}
\endequation

\noindent and  $f_1(m^2_2)$ is arbitrary function.

If in (\ref{10}) $1<Re \ s<2$, then the solution of (\ref{10})
is convergent Feynman integral

\equation
\Gamma _s = \int \mbox{d} ^4 k \frac{1}{ [(p-k)^2 +
m^2_1](k^2)^s}    \label{11}
\endequation

\noindent (This can be proved by using of formula (\ref{7a}).)

Introducing Feynman parameters, one can write $\Gamma _s$ in the
form

\equation
\Gamma _s = \frac{\pi ^2}{s-1} \int _0^1 \mbox{d} x \left[
\frac{x}{(1-x)(xp^2 + m^2_1)} \right]^{s-1} \label{12}
\endequation

We see that $\Gamma _s$ can be analytically continued in the
strip $0<Re \ s <1 $ before the integration with respect to $x$.
Substituting (\ref{12}) in (\ref{9}) and integrating with
respect to $s$, one obtains:

\equation
\Gamma ^{(1)} = - \pi ^2 \int _0^1 \mbox{d} x \ \ln \left( 1+
\frac{(1-x)(xp^2 + m^2_1)}{xm_2^2} \right) \  + \ f_1 (m_2^2)
\label{13}
\endequation

Let $\mu ^2$ is arbitrary constant with dimension of
$\mbox{ [mass] }^2$. Then (\ref{13}) can be rewritten in the
form

\equation
\Gamma ^{(1)} = \left\{ - \pi ^2 \int _0^1 \mbox{d} x \ \ln
\left( \frac {x(1-x)p^2 + xm^2_1 + (1-x)m^2_2}{\mu ^2} \right)
\right\} + \pi ^2 \left( 1+ \ln \frac {\mu ^2}{m_2^2} \right) +
f_1 (m_2^2)   \label{14}
\endequation

Equation (\ref{8}) can be solved in the analogous way. Comparing
results, one can prove that the sum of terms out of braces
in (\ref{14}) is constant. It can be included in the definition
of $\mu ^2$. So, finally, we obtain familiar result:

\eqnarray
\Gamma ^{(1)} (p^2, m_1^2, m_2^2) = - \pi ^2 \int \limits _0^1
\mbox{d} x \ln \left( \frac{x(1-x)p^2 + x m_1^2 +(1-x)m_2^2}{\mu
^2} \right)  \nonumber \\
=\pi ^2 \left\{ \ln \frac{\mu ^2}{m_1 m_2} - \frac{(m_1^2 -
m_2^2)}{2p^2} \ln \frac{m_2^2}{m_1^2} + f \ln \frac{p^2 +m_1^2
+m_2^2 - 2p^2 f}{p^2 +m_1^2 +m_2^2 +2p^2f} \right\} \label{15}
\endeqnarray

\noindent where

$$f= \sqrt {\frac{(m_1^2 -m_2^2)^2}{4p^4} + \frac{m_1^2 +
m_2^2}{2p^2} + \frac{1}{4}}   $$

We see that finite result for divergent diagram on Fig.4 can be
obtained without any regularization and application
of $R$-operation.

Now let us consider less trivial application of our theory. We
will calculate two-loop diagram on Fig.5. In general, this
diagram depends on three different masses $m_1, m_2, m_3$.
For simplicity, we will consider only the most important case $m_1
= m_2 \equiv m, \ m_3 \equiv M $. General case can be treated
analogously.

Power expansions for this diagram were investigated for equal
mass case in \cite{M} and for general case in recent work
\cite{BB}. See also \cite{BT} for corresponding numerical
results.

Up to a polynomial of the first degree with respect to $p^2$, the
corresponding Feynman integral can be defined by equation:

\equation
(\bigtriangleup _{pM})^2 \Gamma^{(2)} (p^2, M^2, m^2) = 16 \pi ^3
\delta
(\vec M) \Gamma ^{(1)} (p^2, m^2, m^2)   \label{17}
\endequation

Up to unessential constant, $\Gamma ^{(1)}$ can be represented
in the following form:

\equation
\Gamma ^{(1)} (p^2, m^2,m^2) = \pi ^2 \int _{4m^2}^{\infty}
\mbox{d} \sigma ^2  \sqrt {1 - \frac{4m^2}{\sigma ^2}} \left(
\frac{1}{p^2 + \sigma ^2} - \frac{1}{\sigma ^2} \right)
\label{18}
\endequation

\noindent Comparing (\ref{7b}), (\ref{17}) and (\ref{18}), we
see that $\Gamma ^{(2)}$ can be represented as following:

\equation
\Gamma ^{(2)} (p^2, M^2, m^2) = \pi ^2 \int _{4m^2}^{\infty}
\mbox{d} \sigma ^2 \sqrt {1 - \frac{4m^2}{\sigma ^2}} \left(
\Gamma ^{(1)} (p^2,M^2, \sigma ^2 ) - \frac{1}{ \sigma ^2}
\varphi (p^2, M^2, \sigma ^2) \right)   \label{19} \endequation

\noindent where $\Gamma ^{(1)} (p^2,M^2, \sigma ^2)$ is defined
by (\ref{15}) and  $\varphi (p^2,M^2, \sigma ^2)$ satisfies the
equation

\equation
(\bigtriangleup _{pm})^2 \varphi = 16 \pi ^3 \delta (\vec M)
\label{20}
\endequation

Using the formula

$$\bigtriangleup _{M} \ln M^2 = 2 \pi \delta (\vec M)  $$

one can represent the solution of (\ref{20}) in the following
form:

\equation
\varphi = \pi ^2 \left\{ M^2 \ln \frac{M^2}{\sigma ^2} + M^2 f_1
(\sigma ^2) + p^2 f_2 (\sigma ^2) + f_3 (\sigma^2) \right\}
\label{21}
\endequation

\noindent where $f_1,f_2,f_3$ are arbitrary functions of $\sigma
^2$. These functions must be defined thus that integral
(\ref{19}) converges. Using explicit formula (\ref{15}) for
$\Gamma ^{(1)} $ , it is easy to prove that one of the possible
choices is

\equation
f_1 =0, \ f_2 = - \frac{1}{2}, \ f_3 = \sigma ^2 \left( \ln
\frac{\mu ^2}{\sigma ^2} -1 \right)  \label{22}
\endequation

\noindent The replacement of functions $ f_1,f_2,f_3$ by any
other ones, for which integral (22) converges,
leads to unessential change of $\Gamma ^{(2)}$ on the polynomial
of the first degree with respect to $p^2$.

Substituting of (\ref{21}) and (\ref{22}) in (\ref{19}), we
obtain our final result:

\equation
\Gamma^{(2)} = \pi ^2 \int _{4m^2}^{\infty } \mbox{d}
\sigma ^2 \sqrt {1- \frac{4m^2}{\sigma ^2} } \left( \Gamma^{(1)}
(p^2,M^2,\sigma ^2) - \pi ^2 \frac{M^2}{\sigma ^2} \ln
\frac{M^2}{\sigma ^2} - \pi ^2 \frac{p^2}{2\sigma ^2} - \pi ^2
\ln \frac{\mu ^2}{\sigma ^2} + \pi ^2 \right)    \label{23}
\endequation

The integrand in (\ref{23}) has the order $O(\sigma ^{-4} \ln
\sigma ^2) $ at $\sigma ^2 \rightarrow \infty $. So integral
(\ref{23}) converges.

To author's knowledge , integral (\ref{23}) can't be expressed
through standard special functions. But integrand in (23) is
rather simple elementary function and so this formula makes
possible to investigate $\Gamma ^{(2)}$ in details. This will be
done in special paper.

One-fold integral representation, that is very similar to
(\ref{23}), was obtained independently in works \cite{BB} and
\cite{BT} by dispersive methods. Analogous representation for
five propogators selfenergy diagram can be found in
\cite{Br1,Br2}.

Now it is unclear, whether our approach to renormalization theory
has principal advantages in comparison with standard
formulation. But, at least, calculations, represented above,
show that our approach gives new effective methods of evaluation
of Feynman integrals. So author believes that proposed formalism
will be useful in various investigations in quantum field
theory.

Author is indebted to D.J.Broadhurst, A.I.Davydychev and
V.A.Smirnov for valuable discussions and comments and also
F.A.Berends, M.Buza, and J.B.Tausk for sending preprints of
their recent works.

\end{document}